\documentclass[conference]{IEEEtran}
\IEEEoverridecommandlockouts
\usepackage{cite}
\usepackage{amsmath,amssymb,amsfonts,fixmath}
\usepackage{algorithm}
\usepackage{subcaption}
\usepackage{graphicx}
\usepackage{textcomp}
\usepackage{xcolor}
\usepackage{ulem}
\usepackage{bm}

\def\BibTeX{{\rm B\kern-.05em{\sc i\kern-.025em b}\kern-.08em
		T\kern-.1667em\lower.7ex\hbox{E}\kern-.125emX}}
\usepackage{pgfplots}
\pgfplotsset{compat=newest}
\usetikzlibrary{plotmarks}
\usetikzlibrary{arrows.meta}
\usepgfplotslibrary{patchplots}
\usepgfplotslibrary{polar}
\usepgfplotslibrary{fillbetween}
\usetikzlibrary{spy,shapes,arrows,positioning,plotmarks,shadows,calc,matrix,fit,patterns}
\usepackage{grffile}
\pgfplotsset{plot coordinates/math parser=false}
\newlength\figureheight
\newlength\figurewidth
\usepackage{tikzscale}
\usepackage{siunitx}
\usepackage{tabularx}
\DeclareSIUnit{\belmilliwatt}{Bm}
\DeclareSIUnit{\dBm}{\deci\belmilliwatt}
\usepackage{standalone,tikz}
\usepackage{booktabs}

\usepackage{algpseudocode}
\algrenewcommand\textproc{}

\begin{document}

\title{LIVE-RIS: A Comprehensive In-Flight Dataset for UAV-Mounted RIS Channel Measurements\\

\thanks{This work was funded by the Deutsche Forschungsgemeinschaft (DFG) under grant MO 1086/17-1, grant SE 1697/22-1, and grant INST 213/1028-1 FUGB. This work was supported in part by the German Federal Ministry of
Education and Research (BMBF) in the course of the 6GEM Research Hub
under grant 16KISK03.}
}

\author{David Müller$^{\ast }$, Kevin Weinberger$^{\dagger }$, Aydin Sezgin$^{\dagger }$, Martin M\"onnigmann$^{\ast }$\\
		$^{\ast }$Automatic Control and Systems Theory, Ruhr-Universit\"at Bochum, Germany,\\
        $^{\dagger }$Institute of Digital Communication Systems, Ruhr-Universit\"at Bochum, Germany,\\
		 \{David.Mueller-r21, Kevin.Weinberger, Aydin.Sezgin, Martin.Moennigmann\}@rub.de}

\maketitle

\begin{abstract}
The dynamic control of the wireless propagation environment enabled by reconfigurable intelligent surfaces (RISs) makes them a promising technology for sixth-generation (6G) wireless networks. Beyond conventional fixed-position deployments, integrating a RIS with an unmanned aerial vehicle (UAV) provides additional spatial flexibility, facilitating favorable Line-of-Sight (LoS) conditions and broadening the range of potential applications.
Although UAV-mounted RIS systems have attracted significant research interest, existing studies rely predominantly on numerical simulations and often overlook the impact of practical constraints and real-world disturbances.
To address this gap, the authors recently presented the first UAV-mounted RIS prototype and experimentally validated its feasibility through real-world experiments.
By using the UAV's extended Kalman Filter (EKF) for real-time RIS reconfiguration, the proposed system effectively mitigates disturbance effects and preserves the performance gains of the RIS-enabled link.
To further advance UAV-mounted RIS research and facilitate broader access to real-world experimental insights, this work provides a comprehensive dataset collected from multiple flight experiments. The dataset includes EKF state estimates, ground-truth measurements, corresponding RIS configurations, and $\mathrm{S}21$ channel data.

\end{abstract}

\section{Introduction}
The advancement of sixth-generation (6G) wireless communication networks has created a growing demand for technologies that can intelligently adapt the wireless propagation environment. In this context, reconfigurable intelligent surfaces (RISs) have emerged as a promising approach, offering the capability to dynamically manipulate electromagnetic waves and thereby enhance signal propagation, communication performance, and network flexibility.
This is enabled by the control of the wavefront, e.g., the phase, amplitude, and frequency, of the impinging signal at each reflecting element ~\cite{Basar2019}. By virtue of these characteristics, the RIS can boost various objectives within the network such as resilience \cite{Weinberger2023} and even facilitate new applications altogether \cite{RISApp}.
To fully exploit its potential, the RIS must be configured according to the requirements of the intended use case.
Besides the configuration, the RIS deployment position also plays a major role, due to the multiplicative path loss caused by the transmitter (Tx)-RIS-receiver (Rx) link.
Satisfying these requirements is particularly challenging in environments with mobile users, blockages, or varying terrain elevations, where favorable propagation conditions cannot be guaranteed.
Unmanned aerial vehicles (UAVs) have been proposed to carry the RIS, such that a line-of-sight (LoS)-link can be guaranteed. 
The benefits of UAV-mounted RIS have been widely studied in the literature.
Their applications extend beyond establishing LoS-links, as they can significantly expand wireless coverage, enhance communication performance and security, and support data collection~\cite{Pogaku2022}.
However, while mounting a RIS on a UAV increases deployment flexibility, it is also exposed to many disturbances, that can degrade channel link quality~\cite{Mueller2024, Mueller2025}.
Despite their practical relevance, these effects have been largely neglected in the literature, as most existing studies rely exclusively on numerical simulations under idealized assumptions.
In our previous work~\cite{Mueller2026}, we developed the first real-time capable UAV-mounted RIS prototype and validated its performance through a series of flight experiments, taking an initial step toward bridging the gap between theoretical studies and practical operation.

In this work, we further advance this effort by providing a comprehensive dataset obtained from an extensive measurement campaign.
The dataset comprises EKF-based state estimates of both UAV and RIS, ground-truth measurements, RIS configurations derived via optimization, and the corresponding $\mathrm{S}21$ channel measurements acquired using a vector network analyzer (VNA).
In addition, we consider multiple Tx, Rx, and UAV-mounted RIS deployment locations, enabling analysis of divers relative geometries.
The dataset can be found on Zenodo\footnote{https://zenodo.org/records/23011126} and GitHub\footnote{https://github.com/DavidMller/risdataset.git}.
 
\section{Prototype Setup}\label{sec:PrototypeSetup}

The UAV-mounted RIS system used in this paper, can be divided into two subsystems: the RIS subsystem and the UAV subsystem. Each subsystem is described in detail in the following subsections.
The communication interface between both subsystems and the data transmission is addressed separately.

\subsection{RIS Subsystem}
\label{sec:RISsetup}

The RIS prototype comprises $M = 120$ elements arranged in a $10\times12$ array spanning an area of $20\times\SI{16}{\centi\meter}$ (see Fig.~\ref{fig:UAVSetup}).
Each element is equipped with a positive intrinsic negative diode to enable 1-bit discrete phase shifting.
The RIS is designed as a binary-switching surface with a \SI{180}{\degree} phase shift at the designed carrier frequency of $\nu=5.385$\si{\giga\hertz}.
During phase shifting, the reflected signal experiences a \SI{3}{\deci\bel} attenuation due to hardware impairments in the phase-shifting circuitry.
We use a Raspberry Pi 4B as the RIS controller, which is connected to the RIS via a serial port with a baud rate of 115200 Bd.
The maximum reconfiguration rate of the RIS, i.e., the frequency at which the phase shifts can be updated, is \SI{50}{\hertz}.

\subsection{UAV Subsystem}
\label{sec:UAV_setup}

\begin{figure}[ht]
	\centering
	\includegraphics[width=1\linewidth] {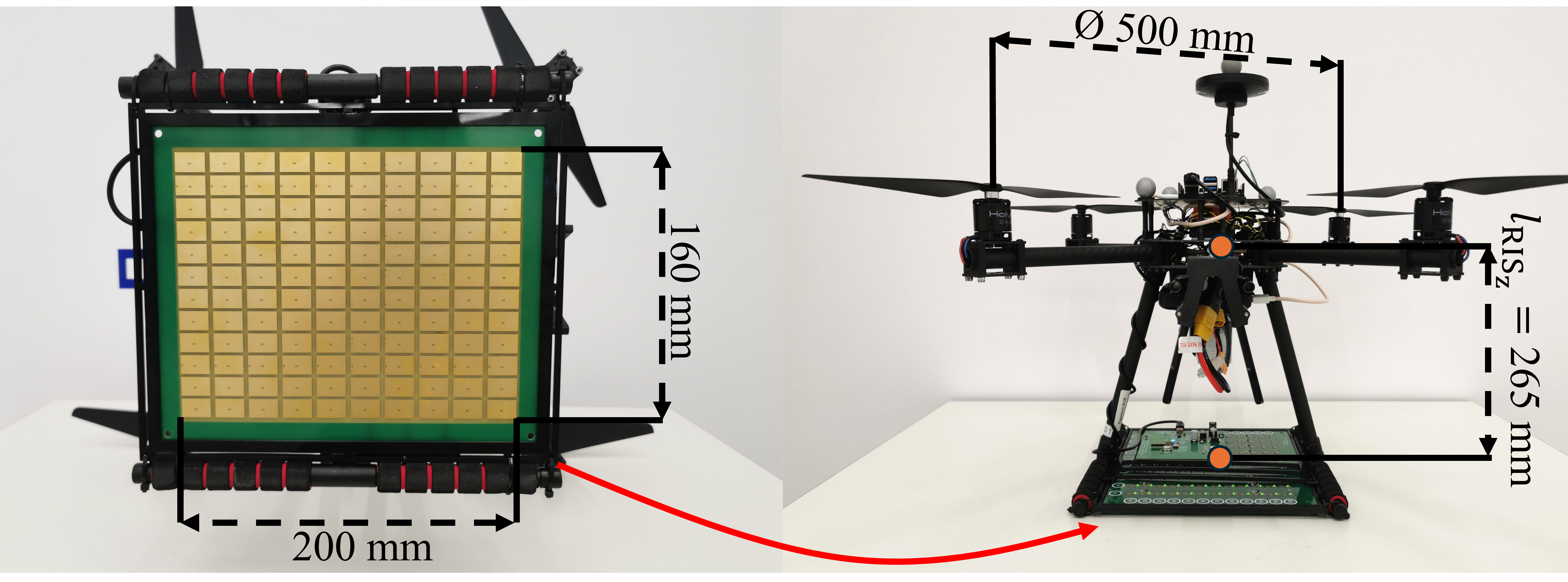}
	\caption{Left: RIS prototype and its dimensions. Right: RIS prototype mounted to a customized Holybro X500. Orange dots represent the origin of the UAV’s body frame (top) and the center of the RIS (bottom).~\cite{Mueller2026}}
	\label{fig:UAVSetup}
\end{figure}

The UAV carrying the RIS is a customized Holybro X500 Quadcopter with a rotor-to-rotor diameter of 500\si{\milli\meter} (see Fig.~\ref{fig:UAVSetup}). It can carry payload of up to 1\si{\kilogram} and is controlled by a Pixhawk Cube Orange flight controller.
The flight controller holds three different inertial measurement units (IMU), where three IMUs are used for redundancy (the primary IMU is a ICM20602, the secondary one a ICM20948 and the third one a ICM20649\footnote{https://ardupilot.org/copter/docs/common-thecubeorange-overview.html}).
The primary and secondary IMUs are mounted on a separate temperature-controlled, vibration-isolated board, while the third IMU is located directly on the flight controller.
The open-source software ArduPilot\footnote{https://ardupilot.org/} (V4.6.3) is installed on the flight controller.
Three different IMUs are installed on the flight controller, as they have different frequency responses and will therefore respond differently to vibrations\footnote{https://docs.cubepilot.org/user-guides/autopilot/the-cube-module-overview}.
We tuned filter parameters for better filtering of vibrations by running an in-flight Fast Fourier Transform on the UAV and analyzing spectra for characteristic frequencies.
Measurements of the IMU are then fused with other sensors by the EKF.
Since the experiments are conducted inside a laboratory, GNSS position information is not available. To enable autonomous indoor flight, we replace the position measurements received by the GNSS antenna with position measurements from a Motion Capture System (MCS).
We use an ESP8266 microcontroller connected to the flight controller running MAVESP8266\footnote{https://github.com/BeyondRobotix/mavesp8266} firmware to send the position information via an additional WiFi connection to the UAV during flight.
The EKF processes the MCS measurements with a frequency of \SI{5}{\hertz}, equivalent to the nominal GNSS update rate.
Details on the MCS are provided in Sec.~\ref{sec:DataCollection}.
We also substitute the on-board barometer and magnetometer measurements with MCS measurements, similar to real-time kinematic GNSS setups~\cite{Mueller2026}.

The RIS is mounted under the UAV, where its center is located $l_{\mathrm{RIS}_z} = 265$\si{\milli\meter} underneath the origin of the UAV body frame (see Fig.~\ref{fig:UAVSetup}), with no offsets in the $x$-$y$-plane.
The Raspberry Pi is mounted on top of the UAV, with almost no vertical offset, and no offsets in the $x$-$y$-plane.

\subsection{Communication between Subsystems}
\label{sec:Communication}

The flight controller of the UAV communicates with the RIS controller through a serial interface with a baud rate of 500000Bd.
Data from the flight controller to the RIS controller is streamed via the MAVLINK 2 protocol\footnote{https://mavlink.io/en/} at a rate of 50\si{\hertz}, matching the sampling frequency of the RIS.
To reduce latency and ensure consistent data transmission, only the most recent UAV EKF data is transmitted.
More specifically, the data consists of the estimated UAV position $p_{k,\mathrm{UAV}}\in \mathbb{R}^{3}$ and the estimated UAV attitude $\Phi_{k,\mathrm{UAV}},\Theta_{k,\mathrm{UAV}},\Psi_{k,\mathrm{UAV}}$ at time step $t_k$, with $k$ denoting the index of the time step. 

\section{RIS Actuation}
\label{sec:RISActuation}

In this section, we give information on how to obtain the RIS pose based on the UAV position and attitude. Subsequently, we introduce the system model assumed in this paper and the optimization approach used to derive the optimal configuration for the RIS.

\subsection{RIS Pose}
\label{sec:Pose}
We first determine the pose of the RIS, which is a prerequisite for the subsequent optimization.
Owing to the rigid connection between the UAV and the RIS, their attitude is identical.
This can be expressed as
\begin{equation}
    \label{eq:rotRis}
    [\Phi_{k,\mathrm{RIS}},\Theta_{k,\mathrm{RIS}},\Psi_{k,\mathrm{RIS}}]
    =[\Phi_{k\mathrm{,UAV}},\Theta_{k\mathrm{,UAV}},\Psi_{k\mathrm{,UAV}}],
\end{equation}
where $\Phi_{k,\mathrm{RIS}}$ denotes the roll angle of the RIS at time step $t_k$.
Pitch and yaw are defined accordingly.
The RIS position $p_{k,\mathrm{RIS}}\in \mathbb{R}^3$ at time step $t_k$ can then be obtained with
\begin{equation}
    \label{eq:p_RIS}
p_{k,\mathrm{RIS}}=p_{k,\mathrm{UAV}} + R(\Psi_{k,\mathrm{RIS}},\Theta_{k,\mathrm{RIS}},\Phi_{k,\mathrm{RIS}})l_{\mathrm{RIS}},
\end{equation}
where $l_{\mathrm{RIS}}=[0,0,-l_{\mathrm{RIS}_z}]^T$ describes the fixed spatial offset between UAV center of mass and the RIS, and $R(\Psi,\Theta,\Phi)\in{\mathbb R}^{3\times 3}$ is 
the product of the rotation matrices 
\begin{equation}\label{rotation_matrix}
    R(\Psi,\Theta,\Phi) = R_z(\Psi)R_y(\Theta)R_x(\Phi). \nonumber
\end{equation}

\subsection{System and Channel Model}
\label{sec:Model}

The system model in this paper assumes a single-antenna Tx transmitting signals towards a single-antenna Rx. It is further assumed that no direct Tx-Rx link exists and a UAV-mounted RIS is used to establish a communication link. The RIS consists of $M$ reflecting elements, each capable of altering the phase of the reflected signal. The reflected signals are then received by the Rx, resulting in an effective channel between Tx-RIS-Rx as
\begin{align}\label{eq:heff}
    h^\mathsf{eff} = \sum_{m=1}^{M} h_m \theta_m g_m, \\[-21pt] \nonumber
\end{align}
where $h_m$ and $g_m$ represent the channel coefficients between
$m$-th RIS element and the Tx/Rx, respectively. Further, the reflect coefficient $\theta_m = A_m(\varphi_m)e^{j\varphi_m}$ is defined by the tunable phase  $\varphi_m\in[0,2\pi)$ induced to the reflected signal at reflecting element $m$ as well as the phase-dependent reflect amplitude $A_m \in [0,1]$.
The cascaded channel coefficient between Tx-RIS-Rx over the \( m \)-th element can consequently be formulated as $h^\mathsf{casc}_m = h_m g_m$.

We determine the cascaded channel components for each reflect element \( m \) based on the geometry of the LoS path, where \( d^h_m \) (\( d^g_m \)) is defined as the distance between the Tx (Rx) and the \( m \)-th RIS element. The cascaded Tx-RIS-Rx channel through the \( m \)-th RIS element is given by
\begin{align}\label{eq:chanModel}
     h^\mathsf{casc}_m = h_m g_m = \left[ \frac{c}{4\pi \nu d^h_m} e^{j\frac{2\pi}{\lambda}d^h_m} \right] \left[ \frac{c}{4\pi \nu d^g_m} e^{j\frac{2\pi}{\lambda}d^g_m} \right],
\end{align}
where \( c \) is the speed of light, \( \nu  \) the carrier frequency, which is given by design of the RIS, and \( \lambda \) the corresponding wavelength.


\subsection{RIS Optimization}
\label{sec:OPtimization}

For optimization of the RIS configuration, we first derive the distances $d^h_m$ and $d^g_m$, i.e., between each reflecting element $m$ for the Tx-RIS and RIS-Rx link, respectively. 
Consequently, the cascaded channel coefficients $h^{\mathrm{casc}}_m$, which is a prerequisite for configuring the RIS optimally, can be determined for every time step $t_k$.
Since this optimization needs to be performed in real time, we need to consider the computational complexity when designing the optimization algorithm.
Due to the blocked path between Tx and Rx, the phase of the effective channel $h^{\mathsf{eff}}$ can be arbitrarily adjusted during the optimization process. 
This can be expressed as
\begin{equation} \label{optForm} 
    \varphi_m^*(C) = C - \varphi'_m, 
\end{equation}
where $C \in \mathbb{R}$ denotes an arbitrary value for the desired phase of $h^{\mathsf{eff}}$ and $\varphi'_m = \frac{2\pi}{\lambda}(d_m^h + d_m^g)$.
We select $C$ from a finite set to ensure sufficiently low computational complexity for real-time operation.
In this paper, we consider the set
\begin{equation}\label{eq:quantizedPhaseShifts}
C\in {\mathcal{C}} = \bigg\{0, \frac{\pi}{4}, \frac{\pi}{2}, \frac{3\pi}{4}, \pi, \frac{5\pi}{4}, \frac{3\pi}{2}, \frac{7\pi}{4}\bigg\}, 
\end{equation}
which results in a suitable trade-off to identify a good solution within the available time~\cite{RIS_proto}.

After optimizing the phase shift $\varphi^*_m$ for every patch $m= \{1, \dots, M\}$, we determine which of the patch state, $\varphi_m^*(C)= 0^\circ$ or $180^\circ$, is the optimal choice. 
We choose a fast rounding operation in light of the real-time requirements.
For the resulting $C\in\mathcal{C}$ candidate configurations, where $C\in\mathcal{C}$ denotes cardinality, we choose the best performing one by assessing and comparing the determined channel quality of the resulting RIS-facilitated links. 
With respect to hardware constraints, this procedure can mathematically be formulated as maximizing $|h_C^{\mathsf{eff}}|$ for all $C\in\mathcal{C}$
\begin{align}
\max_{C\in\mathcal{C}} |h_C^{\mathsf{eff}}| = & \Big|\sum_{m=1}^{M} {h}^{\mathsf{casc}}_m A_m(\text{rd}(\varphi_m^*(C))) \text{e}^{j \text{rd}(\varphi_m^*(C))}\Big|\label{eq:optEq}\\[-8pt] \nonumber
 \text{with}& \,\, A_m(\tau) = \begin{cases} $0.5012 (-3\text{dB})$ , \,\text{if } \tau = \pi \\ 1 , \quad\quad\quad\quad\,\,\, \,\,\,\,\:\text{otherwise}\end{cases}, \\ \nonumber
 &\quad \text{rd}(\tau) = \begin{cases} \pi , \quad\text{if } \frac{\pi}{2} \leq\tau < \frac{3\pi}{2} \\ 0, \quad\text{otherwise}\end{cases}\quad\,\,\,\,\,\,\,  , \\[-1pt] \nonumber
 &\,\,\,\, C\in {\mathcal{C}} = \bigg\{0, \frac{\pi}{4}, \frac{\pi}{2}, \frac{3\pi}{4}, \pi, \frac{5\pi}{4}, \frac{3\pi}{2}, \frac{7\pi}{4}\bigg\}   \\[-13pt] \nonumber
\end{align}
in every time step $t_k$. This optimization can be carried out efficiently, because the optimal $C$ and $\varphi^*_m(C)$ can be found independently for each patch by comparing the ${|\mathcal{C}|}= 8$ choices and rounding subsequently.

\section{Experimental Setup}\label{sec:ExperimentalSetup}

\begin{figure}[t]
	\centering
	\includegraphics[width=0.96\linewidth] {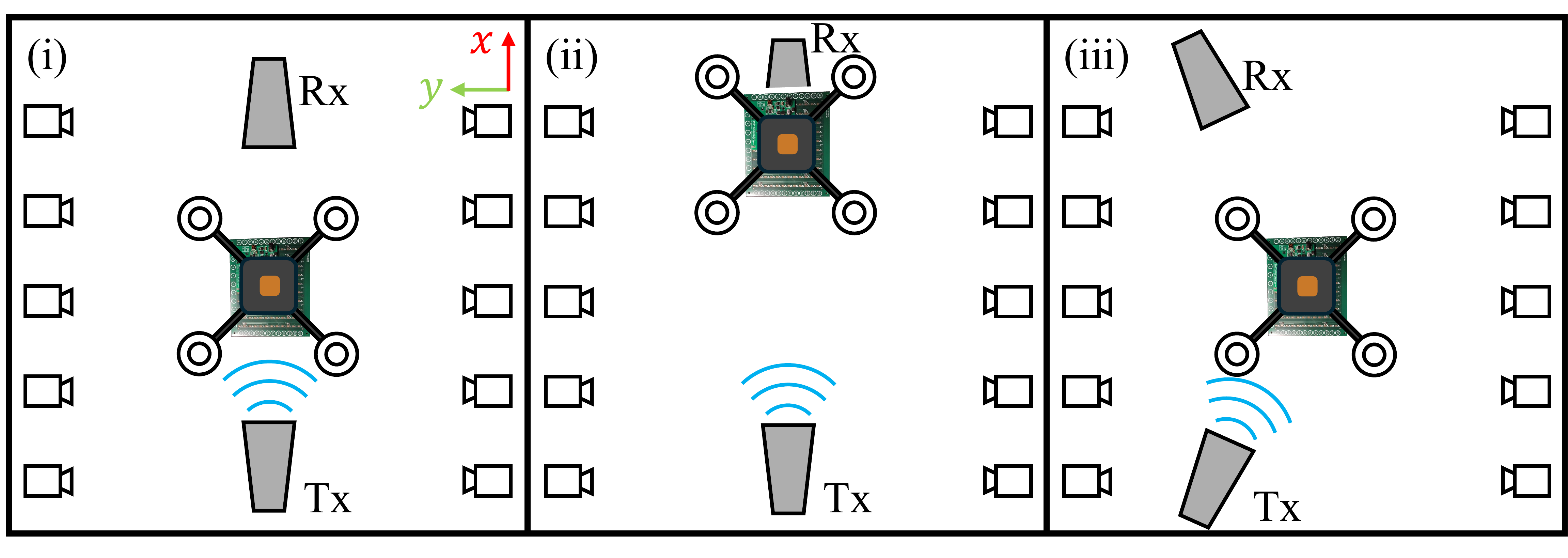}
	\caption{Schematic representation of the three scenarios considered for UAV-mounted RIS performance evaluation. From left to right: (i) RIS positioned at the midpoint between Tx and Rx, (ii) RIS positioned above the Rx, and (iii) RIS positioned off the direct path between Tx and Rx.}
	\label{fig:scenarios}
\end{figure}

In this section, we present the experiment design and the data acquisition process.
Furthermore, we provide details on data processing and synchronization of the collected datasets.
All attitude and position information are provided in the NWU frame, i.e., a right-handed coordinate system with the $x$-axis pointing north, the $y$-axis west, and the $z$-axis upward, as defined in Fig.~\ref{fig:scenarios}.

\subsection{Experiments}
\label{sec:Experiments}

All experiments are conducted in a flight lab under the same environmental conditions.
We consider three scenarios with varying Tx, Rx, and UAV-mounted RIS locations to capture both optimal and non ideal relative geometries.
This accounts for practical deployment conditions, where a UAV-mounted RIS cannot always be positioned optimally with respect to the Tx and Rx. Similarly, three RIS configuration methods are considered to evaluate the benefits and limitations of dynamic, static, and passive reflection.
We treat hover flight in all scenarios, i.e., the UAV is commanded to maintain a fixed target position.
We restart the UAV and calibrate the IMUs on the flight controller before every flight to ensure the same initial conditions.
Each flight has a duration of \SI{40}{\second}, where the first and last \SI{5}{\second} are used for takeoff and landing.
From the remaining \SI{30}{\second}, the first and last \SI{10}{\second} are used to alternate between the optimal RIS configuration and the RIS in its off state.
This procedure is used to synchronize the datasets, hence, the data collected during these intervals are excluded from the dataset.
The remaining \SI{10}{\second} are used to apply the respective RIS configuration and collect measurement data.

\subsubsection{UAV Flight Geometries}

In the first scenario (i), the UAV-mounted RIS is positioned
at the midpoint of the direct link between Tx and Rx, corresponding to the specular reflection path and thus representing an optimal placement for a (non-reconfigurable) reflective surface such as a metal plate.
The target position of the UAV-mounted RIS is set to $(0,0,2)$ and the Tx and Rx are located at $(-1.36,0,0.7)$ and $(1.34,0,0.7)$, respectively.
In the second scenario (ii), the UAV-mounted RIS is positioned directly above the Rx.
Since the RIS-assisted link consists of two cascaded propagation paths, placing the RIS close to either Tx or Rx can reduce the multiplicative path loss.
Here, the target position is set to $(1.2,0,2)$ and the Tx and Rx are located at $(-1.36,0,0.66)$ and $(1.55,0,0.75)$, respectively.
In the third scenario (iii) the UAV-mounted RIS is not deployed along the direct path between Tx and Rx in order to evaluate the impact of a flexible off-axis RIS placement that does not follow the specular reflection path.
Similar to scenario (i), the target position of the UAV-mounted RIS is set to $(0,0,2)$. However, the Tx and Rx are positioned at $(-1.55,0.75,0.7)$ and $(1.65,0.77,0.69)$, respectively.
In every scenario, the horn antennas of the Tx and Rx are oriented towards the target position of the UAV-mounted RIS and aligned with the RIS element polarization.
All scenarios are schematically illustrated in Fig.~\ref{fig:scenarios}.

\subsubsection{RIS Configuration Methods}
We measure the performance of the UAV-mounted RIS using three different RIS configuration methods in each scenario. For each method, two flights are conducted, resulting in a total of six flights per scenario.
The first method (a) continuously updates the RIS according to the optimal phase profile with respect to the channel gain, computed from the most recent UAV EKF estimate using the optimization in \eqref{eq:optEq}. The RIS configuration is updated at a frequency of \SI{50}{\hertz}. This method represents the adaptive case, where the RIS continuously accounts for the changing UAV pose during flight.
The second method (b) applies a single RIS configuration obtained by solving \eqref{eq:optEq} once prior to flight for the respective target position. The configuration is then kept constant throughout the entire flight. This reflects the idealized assumption commonly adopted in theoretical studies, where the RIS is optimized for a fixed UAV position and does not account for position changes during practical flight conditions.
For the third method (c), the RIS is deactivated, i.e., no phase shifts are applied to the reflecting elements. This configuration effectively represents a passive metallic reflector and provides a baseline against which the benefits of RIS-based beamsteering can be evaluated.

\subsection{Data Collection}
\label{sec:DataCollection}

The EKF state estimates of the UAV as well as the corresponding RIS pose and the RIS configurations and the respective EKF timestamp are logged on the Raspberry Pi at at frequency of \SI{50}{\hertz}, equivalent to the RIS update rate.

Reference data for the UAV position and orientation, as well as the data used for GPS substitution (see Sec.~\ref{sec:UAV_setup}), are measured and recorded using an MCS.
Specifically, 10 Vicon Vantage V5 infrared cameras and the software Tracker 3 were installed. This system is able to track objects equipped with an asymmetric pattern of reflective markers with a root mean squared error of less than \SI{0.2}{\milli\meter} with frequencies of up to 420\si{\hertz}\footnote{https://www.vicon.com/}.
We calibrate the MCS once after it has reached its working temperature.
The motion capture system sampling frequency is set to \SI{100}{\hertz}.

To validate the performance of the UAV-mounted RIS during the experiments, we use a VNA of type Keysight P5026B equipped with the S9010B software option, which enables time-gating to isolate the signal component only reflected by the RIS. Specifically, a time gate spanning 61-71\si{\nano\second} is applied, thereby suppressing undesired multipath components and extracting only the RIS-reflected contribution at the target position. The $\mathrm{S21}$ parameter is measured over a bandwidth of \SI{650}{\mega\hertz}, spanning from \SI{5.225}{\giga\hertz} to \SI{5.875}{\giga\hertz}, with 201 frequency points, providing a frequency-resolved characterization of the channel amplitude and phase response. The measurements are conducted in \SI{50}{\hertz} intervals to ensure synchronization with the RIS switching rate, enabling temporally aligned channel acquisition. Two VNA ports serve as the transmitter and receiver, respectively, and are each connected to a directional horn antenna of type LB-187-15-C-SF (A-Info). Within the considered frequency range, the antenna gain is at least \SI{16.35}{\deci\bel i}.
\begin{figure}[t]
	\centering
	\includegraphics[width=0.945\linewidth] {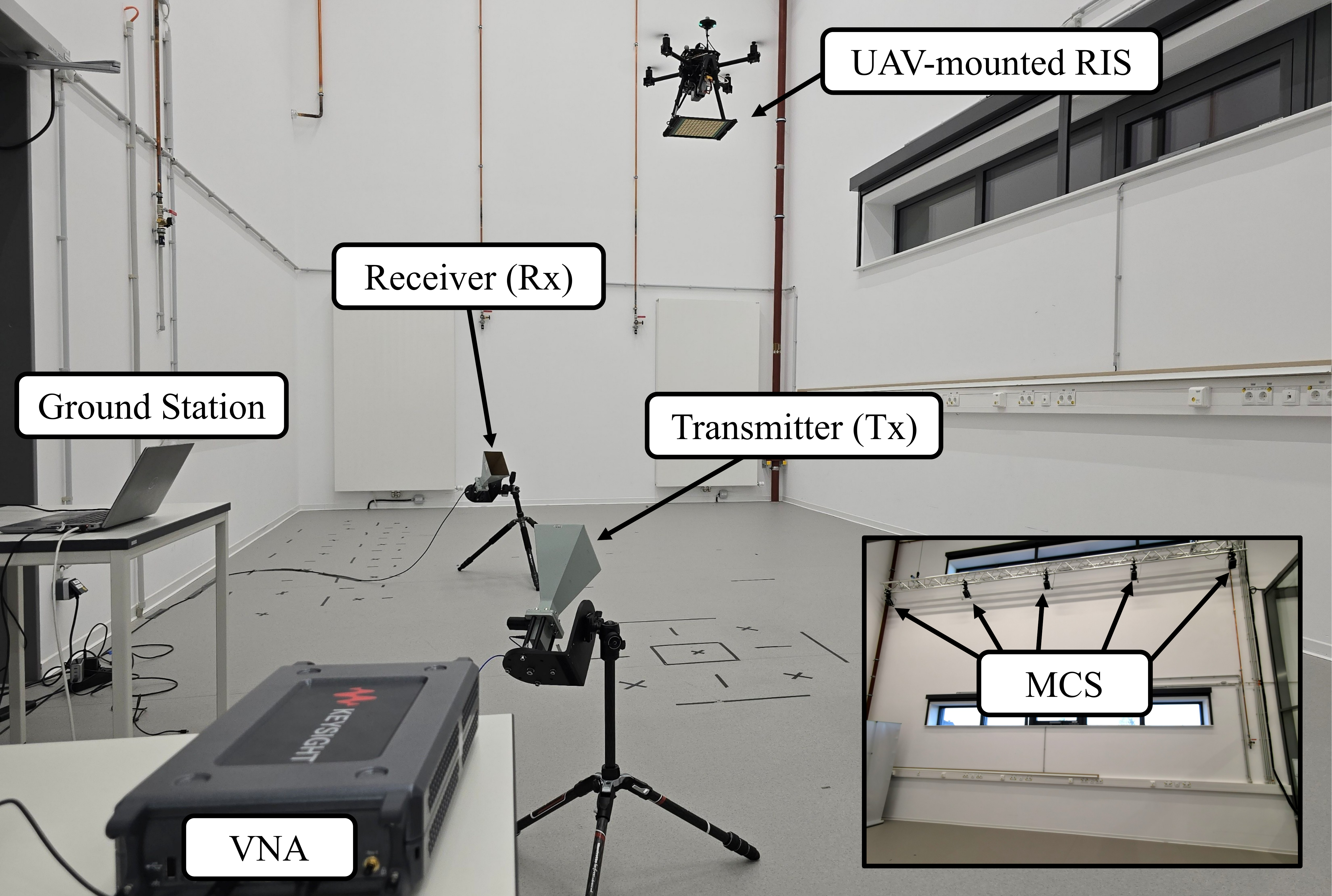}
	\caption{Experimental setup for scenario (iii). Small image in the bottom right corner shows the placement of the motion capture cameras.}
	\label{fig:Experiments}
\end{figure}
The complete setup for the experiment in scenario (iii) is shown in Fig.~\ref{fig:Experiments}.

\subsection{Processing of data}
\label{sec:ProcessingData}

We synchronize the MCS, UAV EKF, and VNA data to remove the temporal misalignment between the sets of data.
All datasets are initially transformed into the NWU coordinate frame.
This is required because UAVs typically operate in the NED frame, whereas the ground truth measurements obtained from the MCS are provided in the NWU frame.
Subsequently, we synchronize the MCS and EKF data by interpolating the MCS measurements to the EKF timestamps.
Since ground truth data is recorded at twice the sampling frequency of the EKF, only the interpolated values at the EKF timestamps are retained.
Consequently, the sets of data are not only synchronized, but the sampling rate of the ground truth data is also reduced to match that of the EKF at \SI{50}{\hertz}.
Based on the synchronized MCS data, which serves as ground truth for the UAV position and orientation, we can compute the true RIS pose according to Sec.~\ref{sec:Pose}.
This enables direct comparison with the estimated RIS pose derived from the most recent UAV EKF estimates, which are recorded on the Raspberry Pi.
Furthermore, the corresponding RIS configuration is already available at the respective timestamps.
Thus, no additional synchronization of the UAV data is required.

Finally we need to synchronize the data measured with the VNA with the RIS pose and configuration.
To this end, we use the alternating phases at the beginning and end of each flight.
By logging the applied RIS configuration on the Raspberry Pi, the changes in signal quality measured by the VNA can be synchronized with the corresponding configurations and their timestamps on the Raspberry Pi, ensuring proper temporal alignment of all datasets.


\section{Dataset Format}
\label{sec:DataFormat}

Each file in the dataset contains timestamped values synchronized according to Sec.~\ref{sec:ProcessingData}. Corresponding timestamps are included in every file. For each \SI{10}{\second} flight segment, the data is sampled at a frequency of \SI{50}{\hertz}, resulting in 500 time steps per file.
Data is provided in the .csv format. 
The dataset is organized hierarchically by scenario, followed by the corresponding RIS configuration, and finally separated into Flight~1 and Flight~2 recordings.
All units are specified in the respective file headers.
Note that in addition to the RIS pose estimates, the dataset also includes the UAV pose estimates together with the corresponding ground-truth measurements to facilitate broader analyses and reproducibility. This enables, for example, the simulation and evaluation of scenarios with larger offsets between the RIS center and the UAV center. 

\begin{figure}[t]
	\centering
	\includegraphics[width=1\linewidth] {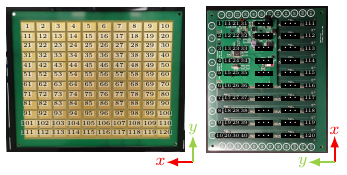}
	\caption{RIS indexing convention utilized in the dataset. The left panel shows the front side, while the right panel shows the top-down view of the downward-facing RIS mounted on the UAV. The illustrated coordinate system follows Fig.~\ref{fig:scenarios}.}
	\label{fig:indexing}
\end{figure}

A special case is the file containing the RIS configuration data. To ensure an unambiguous assignment to each timestamp, the RIS configuration is represented as a binary row vector. Each entry corresponds to an individual RIS element, where a value of 0 indicates that the element is inactive, while a value of 1 denotes that the element applies a \SI{180}{\degree} phase shift to the incoming wavefront. The mapping between the entries of the binary row vector and the corresponding RIS elements is illustrated in Fig.~\ref{fig:indexing}. More precisely, the left panel shows the RIS from the front, while the right panel shows the RIS from the UAV perspective when mounted facing downward. Both panels indicate the corresponding $x$- and $y$-axes of the utilized coordinate system. From the UAV perspective, the longer side of the RIS is aligned with the $x$-axis, and the elements are indexed column-wise from top to bottom, starting at the top-left corner.

\section{Data Validation}
\label{sec:DataValidation}

\begin{figure*}[t]
	\centering
	\includegraphics[width=0.975\linewidth] {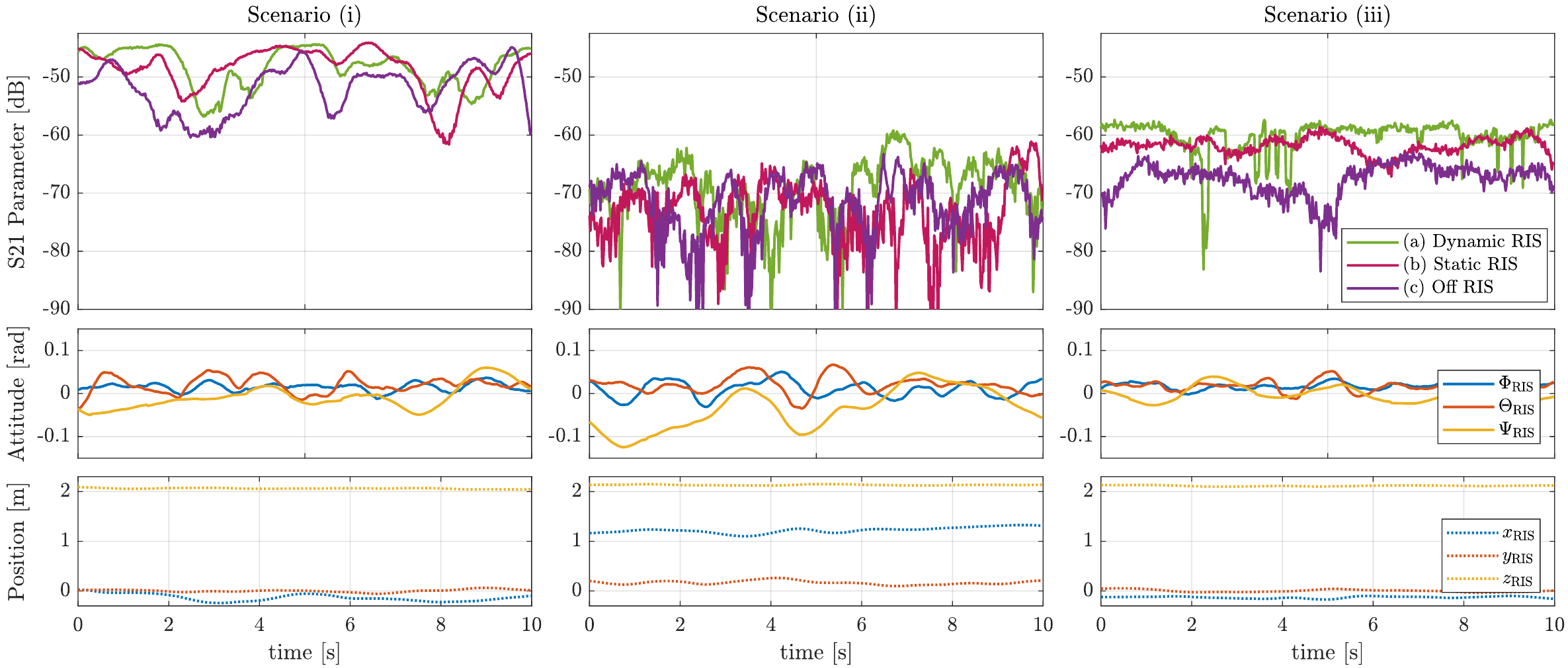}
	\caption{Measured S21 magnitudes for different RIS configurations and deployment locations. First row shows the achieved S21 magnitudes for scenarios (i) - (iii). Green, pink, and purple lines correspond to RIS configuration methods (a), (b), and (c), respectively. Second and third row show representative RIS poses encountered during the flights for method (a).}
	\label{fig:Validation}
\end{figure*}

In this section, we validate the measurement data and provide insights into the information that can be derived from the obtained measurements.
Figure~\ref{fig:Validation} depicts the achievable $\mathrm{S}21$ magnitudes obtained using the three different RIS configuration methods (a) - (c) for scenarios (i) - (iii).
Furthermore, the figure shows the ground-truth (MCS) attitude and center-position of the RIS, only for the flight conducted with the dynamic RIS.
This flight is representative of a typical hover flight.
Since the UAV trajectory is not affected by the RIS configuration, its pose can be considered representative of the other configuration methods as well.
The individual poses for the other methods are therefore omitted, as the small variations between flights would obscure the figure without providing additional insight.

Nevertheless, these small variations can still have a noticeable impact on the measured propagation characteristics, as illustrated in scenario (i), where the UAV-mounted RIS is positioned midway between Tx and Rx.
In this case, the optimal RIS configuration at the target position for configuration method (b) and (c) coincide, leading to identical RIS configurations for the static and off RIS cases.
Despite this, differences of up to \SI{10}{\deci\bel} can be observed in the measured magnitudes, caused by these slight differences in the UAV trajectory between flights.
On average, the corresponding magnitudes differ by approximately \SI{3}{\deci\bel}, emphasizing the impact of inherent motion dynamics on the achievable link performance, a factor that is often neglected in purely theoretical UAV-mounted RIS performance analysis.
When comparing the different scenarios, it becomes evident that the dynamic RIS (a) outperforms both 
static (b) and off RIS (c).
In scenario (i), the dynamic RIS achieves an average gain of approximately \SI{0.5}{\deci\bel} over the static RIS, effectively mitigating disturbances.
For scenario (ii) and (iii), these differences become more pronounced, with the dynamic RIS (a) outperforming the static configuration (b) by up to \SI{2.5}{\deci\bel} on average.
Beyond these average performance gains, the static RIS configuration also exhibits a considerably higher sensitivity to disturbances during flight. In scenario (ii), it is inferior to the RIS in its off state, while outperforming it in scenario (iii). This behavior can be attributed to the UAV position relative to the Tx and Rx. As shown in \cite{Uncertainty_Prop_pos}, positioning the RIS directly above the antennas is particularly sensitive to positional uncertainty, as even small changes in the UAV position can cause pronounced variations in the effective channel. This effect is also reflected in the measured results shown in the figure.
These observations highlight the importance of an real-time adaptable RIS for ensuring stable and reliable signal transmission in dynamic scenarios.

We can derive additional insights from Fig.~\ref{fig:Validation} regarding the influence of the UAV-mounted RIS deployment location on the achievable $\mathrm{S}21$ magnitudes.
When the RIS is positioned midway between Tx and Rx, $\mathrm{S}21$ magnitudes of up to \SI{-48.26}{\deci\bel} are achieved, whereas for the other deployment locations, the magnitudes decrease by more than \SI{20}{\deci\bel}.
Due to the increased sensitivity to positional variations, all configuration methods exhibit worse performance in scenario (ii), where the UAV-mounted RIS is positioned directly above the Rx. Moreover, the comparatively low UAV altitude places the Rx directly beneath the UAV, exposing it to rotor downwash and resulting in less stable flight and faster attitude changes. This is also reflected by the larger attitude variations observed in scenario (ii), which further degrade the RIS link performance.

\section{Conclusion}

We presented a dataset for UAV-mounted RIS operation under real flight conditions.
The dataset comprises synchronized $\mathrm{S}21$ measurements, UAV and RIS pose information obtained from both the UAV's EKF and a MCS, as well as corresponding RIS configurations for multiple deployment scenarios and configuration approaches.
Initial analyses based on the dataset demonstrate that both the RIS configuration approach and the deployment location significantly influence the achievable link performance.
Furthermore, the results reveal that even small disturbances can substantially affect the measured $\mathrm{S}21$ magnitudes, highlighting limitations of purely theoretical analyses.
The presented dataset provides a foundation for future research on UAV-mounted RIS systems and contributes toward bridging the gap between theoretical investigations and practical real-world deployments.


\bibliographystyle{IEEEtran}
\bibliography{./bib} 

\end{document}